Earth, Planets and Space

# Structured regularization based velocity structure estimation in local earthquake tomography for the adaptation to velocity discontinuities

Yohta Yamanaka[1], Sumito Kurata[1*] , Keisuke Yano[2], Fumiyasu Komaki[1], Takahiro Shiina[3] and Aitaro Kato[4]

**Abstract**

We propose a local earthquake tomography method that applies a structured regularization technique to determine sharp changes in Earth's seismic velocity structure using arrival time data of direct waves. Our approach focuses on the ability to better image two common features that are observed in Earth's seismic velocity structure: sharp changes in velocities that correspond to material boundaries, such as the Conrad and Moho discontinuities; and gradual changes in velocity that are associated with pressure and temperature distributions in the crust and mantle. We employ different penalty terms in the vertical and horizontal directions to refine the earthquake tomography. We utilize a vertical-direction (depth) penalty that takes the form of the $l_1$-sum of the $l_2$-norms of the second-order differences of the horizontal units in the vertical direction. This penalty is intended to represent sharp velocity changes caused by discontinuities by creating a piecewise linear depth profile of seismic velocity. We set a horizontal-direction penalty term on the basis of the $l_2$-norm to express gradual velocity tendencies in the horizontal direction, which has been often used in conventional tomography methods. We use a synthetic data set to demonstrate that our method provides significant improvements over velocity structures estimated using conventional methods by obtaining stable estimates of both steep and gradual changes in velocity. We also demonstrate that our proposed method is robust to variations in the amplitude of the velocity jump, the initial velocity model, and the number of observed arrival times, compared with conventional approaches, and verify the adaptability of the proposed method to dipping discontinuities. Furthermore, we apply our proposed method to real seismic data in central Japan and present the potential of our method for detecting velocity discontinuities using the observed arrival times from a small number of local earthquakes.

*Correspondence: kurata@mist.i.u-tokyo.ac.jp
[1] Graduate School of Information Science and Technology, The University of Tokyo, Tokyo, Japan
Full list of author information is available at the end of the article

**Graphical Abstract**

Local earthquake tomography via structured regularization

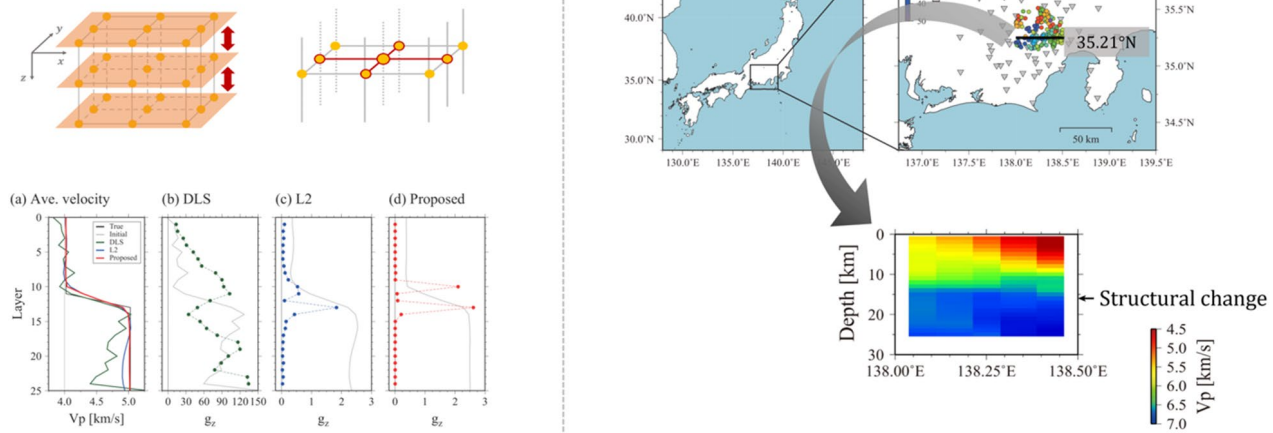

## Introduction

Earthquake tomography methods are used to estimate seismic velocity structure in Earth's crust. The crust is an approximately 10–50-km-thick layer that covers Earth's surface (Bassin 2000), and hosts intense shallow seismicity. Local earthquake tomography (LET) has often been used to capture the high-resolution, three-dimensional (3-D) crustal structure of a given region (e.g., Aki and Lee 1976; Thurber 1993) and to relocate earthquake hypocenters (e.g., Thurber 1983). Therefore, LET and associated approaches provide fundamental information for understanding the mechanisms of earthquake generation in and around the crust (e.g., Alesssandrini et al. 2001; Zhang and Thurber 2003; Nugraha and Mori 2006).

These tomography methods work adequately when a large number of different ray paths are produced by well-distributed source and receiver arrays. However, seismic sources are often localized around fault zones, and most seismic observation stations are deployed near Earth's surface, resulting in an inhomogeneous distribution of seismic ray paths owing to these uneven source and receiver distributions. Therefore, LET commonly suffers from the unstable estimation of structural parameters, or overfitting. Regularization approaches in the inversion have effects on mitigating such instability and overfitting problems. In LET, Laplacian regularization, one of the regularization methods based on $l_2$-norms, has often been used to stabilize crustal seismic velocity structure estimations (e.g., Lees and Crosson 1989; Zhang et al. 1998; Moran et al. 1999). The penalty for dissimilar velocity via $l_2$-norms yields smooth fluctuations in seismic velocity. The smoothed estimates are reasonably acceptable in seismology, because spatial pressure and temperature variations, which are key factors affecting the seismic velocity structure, are generally gradual. However, such regularization often ignores an important component of crustal structures; i.e., a velocity discontinuity that is due to a rapid change in seismic velocity and often represents either a geological boundary or a solid–liquid contact. The Conrad and Mohorovičić (Moho) discontinuities are well-known, and have been incorporated into one-dimensional (1-D) velocity models, such as the PREM (Dziewonski and Anderson 1981) and IASP91 (Kennett and Engdahl 1991) models. Furthermore, there may also be local discontinuities in the crust that are difficult to image, such as the boundary between a sedimentary basin and basement rocks. It is, therefore, desirable to obtain stable estimates of both steep and gradual changes in seismic velocity. One way to overcome the pitfall of $l_2$-norm-based regularization is to place grid points along a discontinuity (Zhao et al. 1992; Moran et al. 1999), but this approach requires accurate prior knowledge of the discontinuity. Another approach is to change grids in adaptation to observations (e.g., Thurber and Eberhart-Phillips 1999), or to assign spatially fine grids enough for image a discontinuity using a dense seismic observation (e.g., Kato et al. 2010, 2021). In our study, we estimate 3-D velocity structure and automatic detection of unidentified velocity discontinuities while fixing the locations of grid points.

To realize reliable estimations of velocity structure including the discontinuities by the framework of tomographic analysis, we develop a new regularization method for 3-D LET that can handle both sharp and gradual changes in the seismic velocity structure of the crust. Specifically, we utilize a combination of the following two penalty terms in a geophysical inverse problem: (i) an $l_1$-sum-type penalty on the $l_2$-norm of the second-order differences between horizontal units in the vertical (depth) direction; and (ii) an $l_2$-sum-type penalty on the first-order differences of the horizontal directions. By combining the two types of penalty term, our proposed method detects steep velocity gradients along the depth direction, with horizontal variations in seismic velocities. In particular, penalty term (i) plays a role to detect unknown structural changes, such as the Conrad discontinuities without prior knowledge.

Penalty term (i) imposed in the vertical (depth) direction is a version of $l_1$ trend filtering (Kim et al. 2009), which is a sparse estimation technique. Sparse estimations with $l_1$-type penalties yield estimates with zero values and work well in balancing the tradeoffs of mitigating overfitting and obtaining estimation accuracy when the estimand has sparse representation (e.g., Tibshirani 1996; Schmidt et al. 2007). Recently, sparse estimations have been utilized in seismology, such as the inference of fault segments (Klinger 2010), the slip distribution of long-term slow slip events (Nakata et al. 2017), and the spatial distribution of changes in seismic scattering properties from small data sets (Hirose et al. 2020). In our case, the vertical-direction penalty causes the distribution of resulting velocities averaged on the horizontal unit to be piecewise linear in the vertical direction; that is, our proposed method enhances sharp structural changes of seismic velocities at depths, where they occur. The horizontal-direction penalty term (ii) produces velocity distributions with smooth fluctuations in the horizontal direction that fit the common understanding of velocity structures in Earth's interior, as horizontal variations in velocity are generally mild compared with vertical variations. We determine all values of hyperparameters (regularization parameters) via cross-validation.

This paper is organized as follows. We first outline the basis of LET and introduce our proposed approach. We then conduct synthetic tests to demonstrate that the proposed method works better than conventional methods in estimating velocity structures with sharp vertical changes. In addition, we apply the proposed method to real seismic data. Results of the analysis indicate the ability of our LET method with structural regularization to clarify structural discontinuities in the crust with 3-D velocity structure, even when the number of available observational data is not large. Additional details on the mathematical formulations and numerical experiments are described in the Appendix.

## Mathematical formulation

In this section, we provide the LET mathematical formulation, focusing on estimations of 3-D velocity structures. We focus only on cases using compressional-wave (P-wave) arrivals, as the description does not depend on a specific seismic phase.

### LET fundamental framework

We first design 3-D grid points to model subsurface velocity structures. Let $v_{x,y,z}$ be a seismic velocity parameter at grid point $(x, y, z)$. In this paper, the $z$ axis indicates depth, and the $x−y$ plane indicates horizontal location. We assume that the grid points are arranged at uniform intervals in the horizontal and vertical directions, respectively. Hereafter, we refer to a plane consisting of grid points that are located in the same depth ($z$) as a "layer". We then calculate the velocity $V_{\tilde{x},\tilde{y},\tilde{z}}$ at an arbitrary point $(\tilde{x}, \tilde{y}, \tilde{z})$ by linear interpolation using the values of the velocity parameters at the nearest eight grid points. This point $(\tilde{x}, \tilde{y}, \tilde{z})$ is not necessarily included in the set of grid points $\nu = \{v_{x,y,z}\}$.

An arrival time contains information on the following factors: the origin time $\tau_i$ of earthquake $i$; the hypocenter location $h_i$ of earthquake $i$; the velocity parameter $v_{x,y,z}$ at grid point $(x, y, z)$; and the ray path from $h_i$ to seismic station $s_j$. These factors are combined using ray theory to provide the predicted arrival time $T_{i,j}^{(\text{cal})}$ from hypocenter $h_i$ to seismic station $s_j$:

$$T_{i,j}^{(\text{cal})} = \tau_i + \int_{h_i}^{s_j} \frac{1}{V_{\tilde{x},\tilde{y},\tilde{z}}} d\rho,$$

where $d\rho$ denotes the element of the path length. Estimations of velocity parameters in LET are usually conducted based on the (damped or regularized) least-squares method. The objective function to be minimized is an additive form of the residual sum of squares (RSS) between the calculated arrival times $T_{i,j}^{(\text{cal})}$ and the observed arrival times $T_{i,j}^{(\text{obs})}$ for all of the available earthquake–station pairs, and penalty terms:

$$\tilde{f} = \sum_{i \in I} \sum_{j \in J} \left( T_{i,j}^{(\text{obs})} - T_{i,j}^{(\text{cal})} \right)^2 + D(\nu, h) + P(\nu), \tag{1}$$

where $h$ is the set consisting of the hypocenter locations and origin times, and $I$ and $J$ are the observed earthquakes and available observation stations, respectively. The second and third terms ($D(\nu, h)$ and $P(\nu)$) represent the penalties on $(\nu, h)$ and $\nu$, respectively. Details of these terms are explained in the next paragraph. After setting the initial values of model parameters, $D(\nu, h)$ and $P(\nu)$, we obtain estimates of objective factors (velocity and hypocentral parameters and ray paths) in an iterative computation. We obtain estimates of objective factors (velocity and hypocentral parameters) by iteratively updating them. In each iteration, velocity and hypocentral parameters are updated jointly using ray path reevaluating for every earthquake–station pair. This update procedure is conducted until the desired accuracy of the tomography is achieved.

The second term, $D(\nu, h)$, in Eq. (1) is the damping term that consists of square norm of model parameter change from the initial values. In the field of earthquake tomography, such estimation with damping term (damped least-square method, DLS) has been utilized (e.g., Aki and Lee 1976; Thurber 1983). Although such damping term has effects in avoiding unstable estimation and overfitting, it generally does not take the spatial information of grid points into account. The third term, $P(\nu)$, in Eq. (1) represents an additional penalty that depends on only velocity parameters and incorporates the spatial information. In the DLS, $P(\nu)$ is not employed. For $P(\nu)$, regularization based on the $l_2$-smoothness has often been used (e.g., Lees and Crosson 1989; Zhang et al. 1998). For example, the following terms can be employed as $P(\nu)$:

$$\lambda_1 \sum_{x,y,z} \sum_{(x',y',z') \sim (x,y,z)} \big(\nu_{x,y,z} - \nu_{x',y',z'}\big)^2,$$
$$\lambda_2 \sum_{x,y,z} ||\Delta \nu_{x,y,z}||_2^2,$$

where the relation $(x',y',z') \sim (x,y,z)$ means that the two grid points are adjacent to each other, $\Delta$ indicates the Laplacian operator, and $||\cdot||_2$ represents the $l_2$-norm. $\lambda_1$ and $\lambda_2$ are non-negative regularization parameters. The former term penalizes dissimilarity among adjacent grid points, and the latter term shrinks the variation in velocity gradients in the three directions. By employing the above penalty terms as $P(\nu)$, we can smooth fluctuations in velocity parameters, as well as suppress the destabilization of estimated parameters. Such regularization methods mitigate overfitting to some extent, yet it often discards the presence of discontinuities, since the resulting estimate smooths discontinuities out. Thus, we propose yet another penalty as $P(\nu)$, which is described in the next subsection, for obtaining more accurate estimations of 3-D velocity structures involving discontinuities using the LET framework.

## Proposed method: structured regularization for 3-D LET

Here we propose a structured regularization approach to accurately image two different velocity changes: sharp velocity changes in the vertical direction at discontinuities; and relatively gradual velocity changes in the horizontal directions. Our objective function, which is minimized to estimate the optimal model parameters, has the form introduced in Eq. (1) employing two additional penalty terms (the vertical-direction regularization term $\Omega_{\text{ver}}$ and the horizontal-direction regularization term $\Omega_{\text{hor}}$) as follows:

$$P(\nu) = \lambda_{\text{ver}} \Omega_{\text{ver}}(\nu) + \tfrac{1}{2} \lambda_{\text{hor}} \Omega_{\text{hor}}(\nu), \qquad (2)$$

where $\lambda_{\text{ver}}$ and $\lambda_{\text{hor}}$ are non-negative regularization parameters. We multiply $\Omega_{\text{hor}}(\nu)$ by $1/2$ for the convenience of computation (also see the Appendix). We obtain estimates of velocity parameters by applying iterative calculations based on the alternating direction method of multipliers (ADMM, Glowinski and Marroco 1975; Gabay and Mercier 1976) to this nonlinear and nonconvex problem. The detailed estimation procedure is described in the Appendix.

The vertical penalty $\Omega_{\text{ver}}$ takes the form:

$$\Omega_{\text{ver}}(\nu) = \sum_z \sqrt{g_z(\nu)}, \qquad (3)$$

$$g_z(\nu) = \sum_{x,y} \big\{ \big(\nu_{x,y,z-1} - \nu_{x,y,z}\big) - \big(\nu_{x,y,z} - \nu_{x,y,z+1}\big) \big\}^2. \qquad (4)$$

The vertical-direction penalty $\Omega_{\text{ver}}$ is the sum of the square root of $g_z$, that is, the $l_2$-norm of the second-order differences between the horizontal layers at different depths. This form is a version of $l_1$ trend filtering (Kim et al. 2009) that has been applied in various research fields (e.g., Tibshirani 2014; Selvin et al. 2016; Guntuboyina et al. 2020). This method is known to be suitable for capturing underlying piecewise linear trends. A notable advantage of $l_1$ trend filtering is a reduction in the penalized elements to zero, in contrast to $l_2$-type regularization, which does not provide this reduction (e.g., Wang et al. 2016). In this study, we utilize this approach to detect and adapt to velocity discontinuities by focusing on suppression of the variation in velocity gradients. Values of $g_z$, the penalized elements, become to be zero when gradient of the average velocities among the $z-1$, $z$, and $z+1$ th layers of depth is constant, and thus $\Omega_{\text{ver}}$ forces the minimizer of Eq. (2) piecewise linearly in the vertical direction. In general, seismic velocity changes sharply around material boundaries, such as the Conrad and Moho discontinuities. Our penalty term enhances the depths at which such sharp changes occur, by detecting the change points of the velocity gradient.

Next, the horizontal penalty $\Omega_{\text{hor}}$ is given by

$$\Omega_{\text{hor}}(\nu) = \sum_{x,y,z} \sum_{(x',y',z) \sim (x,y,z)} \big(\nu_{x,y,z} - \nu_{x',y',z}\big)^2. \qquad (5)$$

The horizontal-direction penalty builds upon the first-order velocity differences between adjacent grid points at the same depth. The term $\Omega_{\mathrm{hor}}$ is the $l_2$-type penalty, which allows the resultant velocity parameters to vary smoothly. The penalty terms in Eqs. (3) and (5) need to be divided by the corresponding grid intervals if the grid points are not arranged at respective uniform intervals.

Figure 1 illustrates how the penalty terms work in the vertical and horizontal directions. Our proposed vertical-direction penalty $\Omega_{\mathrm{ver}}$ is based on the $l_1$-sum of $l_2$-norm (sum of the square root of $g_z$), which suppresses variations in the average–velocity gradient in the vertical direction. Using the proposed approach, we can adapt to sharp velocity changes due to geological discontinuities at depth, and there is no requirement for prior information on the location of the discontinuity. It is noted that regularization parameters $\lambda_{\mathrm{ver}} > 0$ and $\lambda_{\mathrm{hor}} > 0$ in Eq. (2) control the smoothness of the resulting velocity structure in the vertical and horizontal directions, respectively. When $\lambda_{\mathrm{ver}}$ is large, velocity gradients are strongly suppressed, and the resulting depth–averaged velocity, therefore, tends to have few steep velocity gradients. In contrast, when $\lambda_{\mathrm{ver}}$ is close to zero, the resulting depth–averaged velocity becomes unsmooth, since the variations of velocity gradients are hardly taken into account. In the horizontal direction, large $\lambda_{\mathrm{hor}}$ tends to make estimated velocity parameters uniform in each layer, and small $\lambda_{\mathrm{hor}}$ permits unsmooth variations. If both $\lambda_{\mathrm{ver}}$ and $\lambda_{\mathrm{hor}}$ are close to zero, the proposed estimation method is almost identical to the DLS method.

## Numerical experiment

We evaluate the performances of our proposed regularization method and two conventional methods—the DLS method and regularization via $l_2$-norm-based smoothing—to determine the effectiveness of our proposed method in reproducing the seismic velocity structure of a given region. The additional penalty term $P(\nu)$ in Eq. (1) is zero when we estimate parameters via DLS. Smoothing methods based on $l_2$-norm have often been used in LET, as mentioned in the Introduction. For the $l_2$-norm-based smoothing, in this experiment we employed the following $l_2$-norm-based penalty term as $P(\nu)$ in Eq. (1):

$$P^{l_2}(\nu) = \lambda_{\mathrm{ver}}^{l_2}\Omega_{\mathrm{ver}}^{l_2}(\nu) + \tfrac{1}{2}\lambda_{\mathrm{hor}}^{l_2}\Omega_{\mathrm{hor}}^{l_2}(\nu), \qquad (6)$$

where

$$\begin{aligned}\Omega_{\mathrm{ver}}^{l_2}(\nu) &= \sum_z g_z(\nu)\\ &= \sum_{x,y,z}\left\{\left(\nu_{x,y,z-1}-\nu_{x,y,z}\right)-\left(\nu_{x,y,z}-\nu_{x,y,z+1}\right)\right\}^2,\end{aligned} \qquad (7)$$

$$\Omega_{\mathrm{hor}}^{l_2}(\nu) = \Omega_{\mathrm{hor}}(\nu),$$

and $\left(\lambda_{\mathrm{ver}}^{l_2}, \lambda_{\mathrm{hor}}^{l_2}\right)$ are non-negative regularization parameters. Both this and our proposed method impose the same penalty in the horizontal direction, because we focus on investigating the advantage of the sparse estimation on the accuracy. We hereafter refer to this method as "$l_2$-smoothness regularization" (or "L2") for notational simplicity. A key difference between $l_2$-smoothness regularization and our proposed methods is the employed norm in the penalties for the velocity structure; the former uses the $l_2$-sum of the $l_2$-norm (sum of $g_z$; Eq. (7)), and the latter uses the $l_1$-sum of the $l_2$-norm (sum of square roots of $g_z$; Eq. (3)). In this experiment, we used the same procedures for the different estimation methods, except for the estimation of velocity parameters for comparing the accuracy of imaging of velocity structures. We applied the algorithm in SIMULPS12 (Thurber 1993),

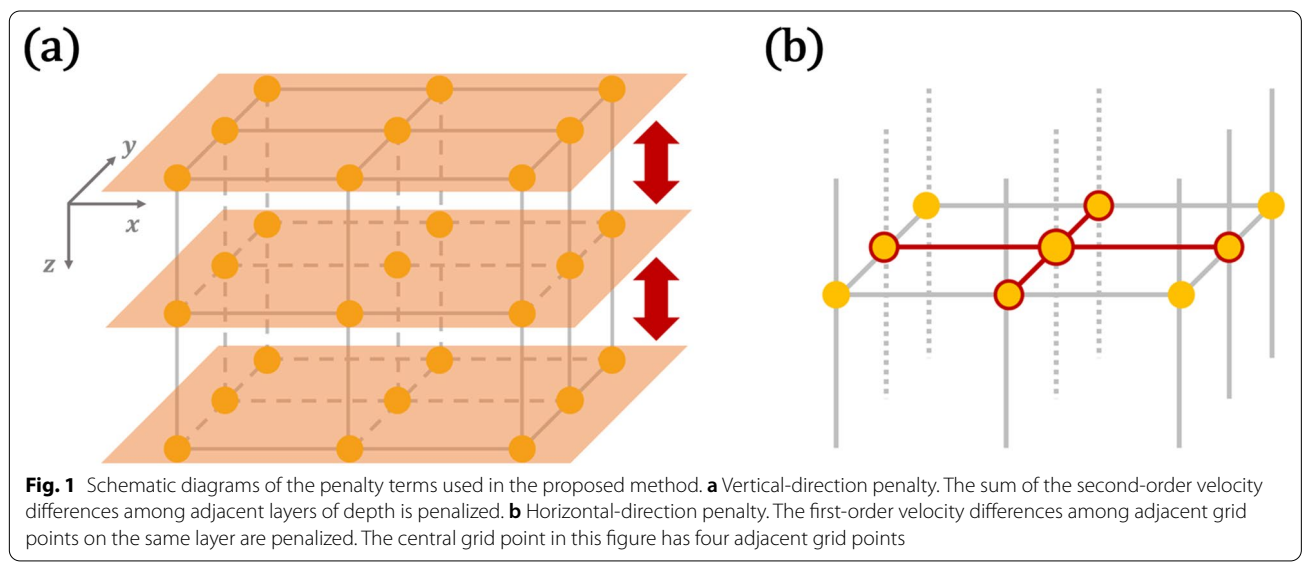

**Fig. 1** Schematic diagrams of the penalty terms used in the proposed method. **a** Vertical-direction penalty. The sum of the second-order velocity differences among adjacent layers of depth is penalized. **b** Horizontal-direction penalty. The first-order velocity differences among adjacent grid points on the same layer are penalized. The central grid point in this figure has four adjacent grid points

a LET-based software package, to determine the hypocenter locations and perform the 3-D ray-tracing calculations through the resultant velocity model for each method.

We determined the regularization parameters via cross-validation. We first split the data set into training and validation data sets. We then estimated the velocity parameters using given regularization parameter values and the training data set comprising a set of prepared values. Finally, we selected the regularization parameter values within the set that minimized the root mean square error (RMSE) of the predicted arrival times for the validation data set.

### Synthetic data

We conducted synthetic tests using the Japan Meteorological Agency (JMA) unified earthquake catalog to investigate the performances of the approaches with different regularizations. The location of the study area is shown in Fig. 2. The data set consists of 199 earthquakes that occurred in central Japan. We obtained 3954 P-wave arrival times from 68 seismic stations in the target region. The arrival times were divided into training and validation data sets, with 2965 arrival times for estimating the velocity parameters and 989 arrival times for validating method accuracy.

We constructed a 26-layer model that extended from 0.0 km (surface) to 25.0 km depth at 1.0-km intervals. We denote the surface layer as "Layer 0" and the layers with grid points at $d$ km depth as "Layer $d$". We then placed 36 ($6 \times 6$) horizontal-directed grid points at an 8.0-km horizontal interval in each layer, with the center of the grid points at 35.25° N, 138.25° E. We set outer points, which surrounded the main target region and have fixed velocity, because some of the hypocenters and stations were located outside the target region. We arranged the outer points as those that were 220 km from the end grid points of each layer in the horizontal direction, and set the "outer layer" at 200 km depth in the vertical direction, to suppress the influence of the velocities at the outer grid points.

We calculated the synthetic P-wave arrival times as follows. We first defined the baseline velocity of each layer, as shown in Fig. 3a. We assumed a 1-D velocity model with a sharp increase in velocity at around Layer 12. We then generated "true" velocities at the grid points by adding ±5% anomalies to the baseline velocities to produce a checkerboard pattern for each layer, as shown in Fig. 3b. Finally, we calculated synthetic arrival times for the available earthquake–station pairs using the true 3-D velocity structure, and generated additional time by adding Gaussian noise with a standard deviation of 0.1 s.

### Results

Figure 4 shows the average velocity–depth profile for each method. When averaging, we used the estimated values of velocity parameters, except for the outer grid points. The initial value of velocity parameter in each grid point was assumed to be 4.0 km/s in this synthetic test. The following regularization parameter values were determined via cross-validation: $\left(\lambda_{\mathrm{ver}}^{l_2}, \lambda_{\mathrm{hor}}^{l_2}\right) = (0.50, 0.06)$ in the $l_2$-smoothness regularization (Eq. (6)), and $(\lambda_{\mathrm{ver}}, \lambda_{\mathrm{hor}}) = (0.10, 0.10)$ in the proposed method (Eq. (2)). The averaged velocities obtained via the estimation methods all capture the sharp increases in velocity around Layer 12, as shown in Fig. 4a. However, the DLS method output shows obvious fluctuations in its estimated velocity structure. This may be due to the 1 km grid interval in the vertical direction being finer than the grid interval that LET studies have generally employed when using data from the nationwide seismic network in Japan (e.g., Matsubara et al. 2017). These unstable DLS estimations indicate that it is difficult to adapt to sharp changes in the velocity structure and avoid ill-posed estimations of the velocity structure without using information on the spatial arrangement of the grid points. $l_2$

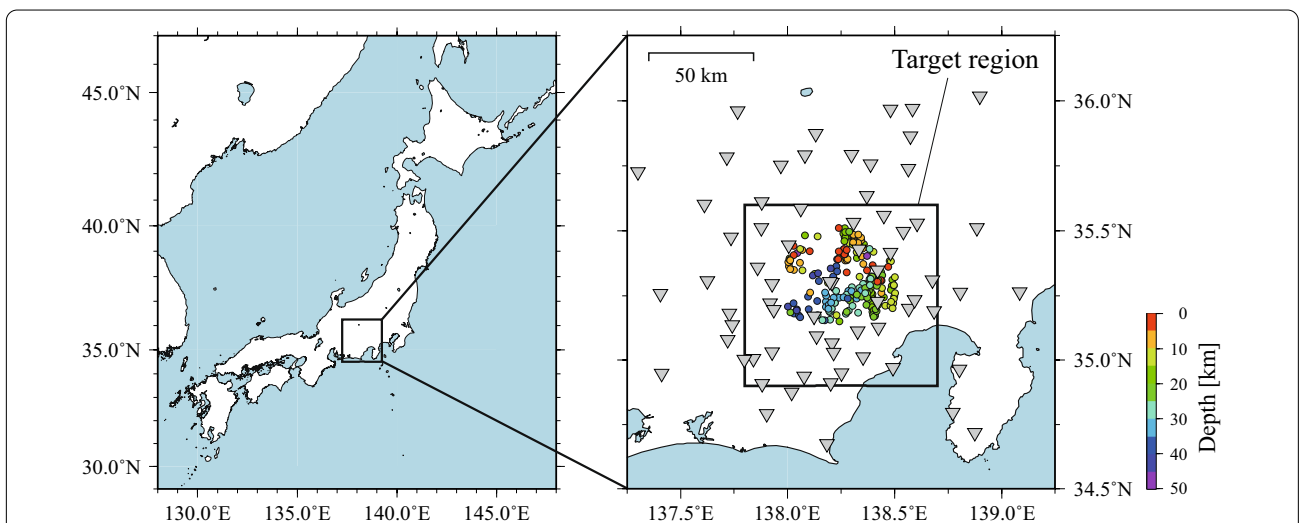

**Fig. 2** (**Left**) Map of the target region (bold rectangle) in Japan. (**Right**) Distributions of epicenters and observation stations that were used in our synthetic test. Circles represent earthquake epicenters, and grey inverted triangles represent observation stations. Epicenters are color-coded by depth

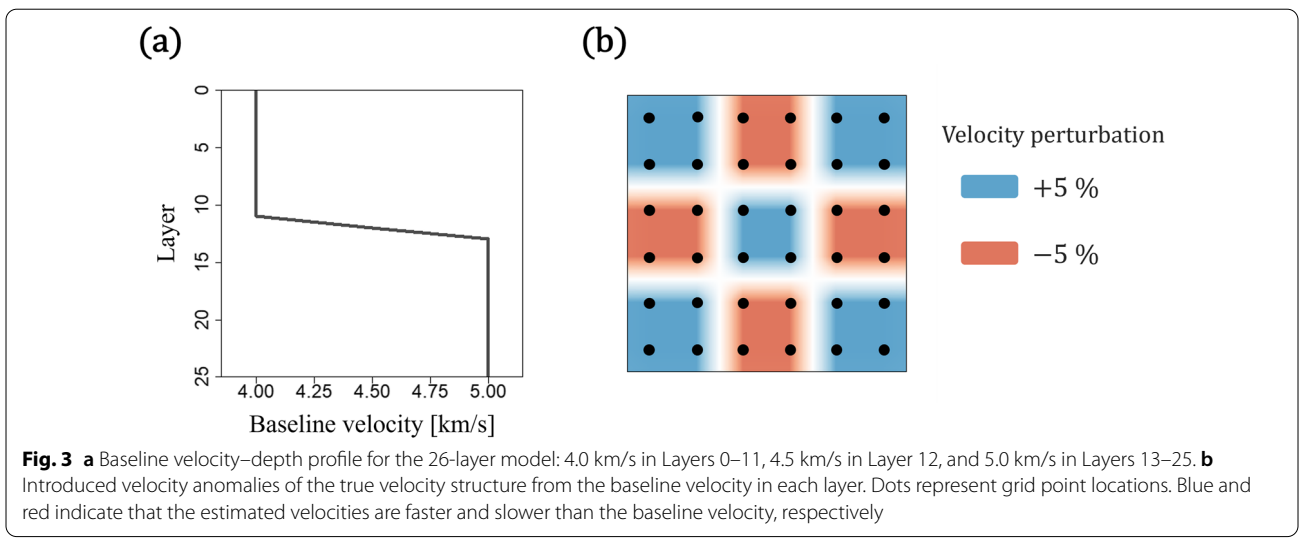

**Fig. 3** **a** Baseline velocity–depth profile for the 26-layer model: 4.0 km/s in Layers 0–11, 4.5 km/s in Layer 12, and 5.0 km/s in Layers 13–25. **b** Introduced velocity anomalies of the true velocity structure from the baseline velocity in each layer. Dots represent grid point locations. Blue and red indicate that the estimated velocities are faster and slower than the baseline velocity, respectively

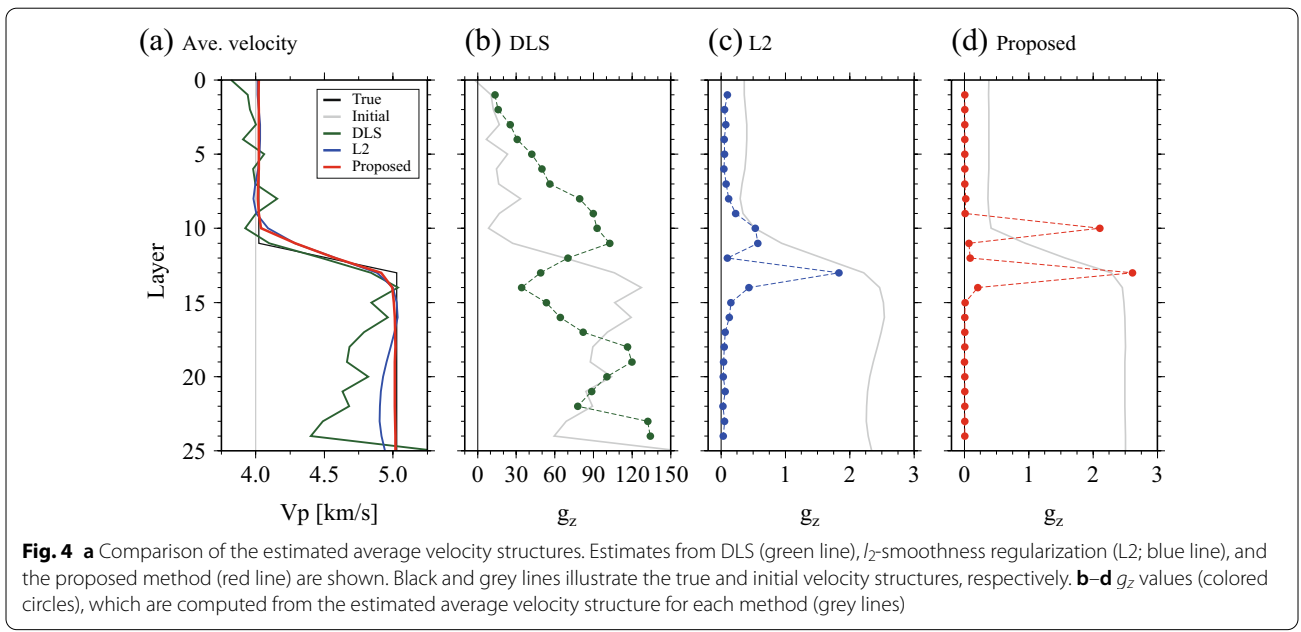

**Fig. 4** **a** Comparison of the estimated average velocity structures. Estimates from DLS (green line), $l_2$-smoothness regularization (L2; blue line), and the proposed method (red line) are shown. Black and grey lines illustrate the true and initial velocity structures, respectively. **b**–**d** $g_z$ values (colored circles), which are computed from the estimated average velocity structure for each method (grey lines)

-smoothness regularization outperformed DLS due to the employed regularization, which reduced the fluctuations in averaged velocities. However, the $l_2$-smoothness regularization estimates at grid points in the layers near and below the velocity jump (Layers 13–25) were unable to reproduce the true velocity structure. Conversely, our proposed method recovered the true average velocities reasonably well, including the sharp increase in velocity around Layer 12. We quantitatively compared the accuracy of estimation of each method by calculating the mean absolute error (MAE):

$$\text{MAE} = \frac{1}{N_g} \sum_{x,y,z} \left| v_{x,y,z}^{(\text{estimates})} - v_{x,y,z}^{(\text{true})} \right|,$$

where $N_g$ is the number of grid points, and $v_{x,y,z}^{(\text{estimates})}$ and $v_{x,y,z}^{(\text{true})}$ are the estimated and true velocity parameters at each grid point $(x, y, z)$, respectively. The values of MAE of DLS, $l_2$-smoothness regularization, and our proposed methods were 0.383, 0.080, and 0.040, respectively.

The norm of the penalty in the vertical direction differs between the $l_2$-smoothness regularization and the proposed method, as shown in Eqs. (3) and (7). The $l_2$-smoothness regularization employs the $l_2$-sum of the $l_2$-norm as the vertical-direction penalty, whereas the proposed method employs the $l_1$-sum of the $l_2$-norm. The $g_z$ values (Eq. (4)), which were evaluated using the obtained velocity structure for each method, are illustrated in Fig. 4b. The $g_z$ values should be zero for most of the layers, except for those around Layer 12, where there is a sudden velocity change, because the true velocity structure was generated from only three baseline velocities (Fig. 3a): 4.0 km/s in Layers 0–11, 4.5 km/s in Layer 12, and 5.0 km/s in Layers 13–25. Most of the computed $g_z$ values for the DLS-estimated velocity structure were far from zero, as shown in Fig. 4b. Although the $l_2$-smoothness regularization-estimated $g_z$ values were relatively small compared with the DLS-estimated values, the penalty terms of $l_2$-smoothness regularization did not reduce $g_z$ to zero. In contrast, most of the $g_z$ values estimated by our proposed method were almost exactly zero, as the penalty terms in this method produce a piecewise linear velocity structure.

We now focus on spatial variations in the estimated velocities in the horizontal units. The checkerboard anomalies imposed on the true velocity structure and the estimated velocity perturbations, both of which were computed from the baseline velocities at each grid point in Layers 1, 12, 20, and 25, are shown in Fig. 5a. DLS tends to estimate amplitude anomalies that are more than 5% smaller/larger than the assumed true structure in this experiment. Both the $l_2$-smoothness regularization and proposed method reproduced the checkerboard pattern in the shallower layers (Layers 0–10). However, $l_2$-smoothness regularization failed to reproduce the assumed checkerboard pattern in the deeper layers (Layers 18–25), whereas the proposed method successfully restored the true structure in most areas (see "Layer 20" and "Layer 25" in Fig. 5a). These results suggest that we can also improve the estimated accuracy about the horizontal-direction variations by grasping the vertical-direction structural changes via the sparse regularization term. Note that, as the spatial locations of hypocenters and stations are non-uniform in the target region (Fig. 2), the number of ray paths differs according to location (e.g., there are relatively few hypocenters in the south part of the target region). Nevertheless, the proposed method succeeded to recover the true structure from the spatially biased data.

Figure 5b illustrates the initial and relocated hypocenters when using our proposed velocity estimation as the velocity estimation. The mean, median, and standard deviation of the estimated errors were 1.48, 1.25, and 1.15 km, respectively. When using the conventional methods for velocity estimation, we obtained similar relocating results: the means of estimated errors using the DLS and $l_2$-smoothness regularization methods were 1.63 and 1.49 km, respectively. These results suggest that our velocity estimation with structured regularization does not influence hypocenter determination. In addition, we compared the proposed method with other regularization methods (Laplacian regularization and other sparse regularization methods via $l_1$-norm) through several experiments, as detailed in the Appendix.

## Discussion

### Size of the velocity jump at a discontinuity

We examined the sensitivity of the three estimation methods (DLS, $l_2$-smoothness regularization, and our proposed method) to the amplitude of a velocity jump in the vertical direction. It is expected that estimation accuracy of the velocity parameters will deteriorate as the size of the velocity jump becomes larger. The initial value of velocity parameter in each grid point was the same as the main experiment in the previous section (uniform velocity of 4.0 km/s). Figure 6 shows the averages of the true and estimated velocities at each layer, and the MAEs for each tested velocity jump. The results of this sensitivity test are shown in Figs. 6a–c and 4 (the case for which the size of the velocity jump is 1.0 km/s). All methods yielded comparable estimation accuracies when there was no velocity jump (constant velocity with depth; Fig. 6a). We also found that the performance of the DLS method degraded gradually as the size of the velocity jump increased. $l_2$-smoothness regularization performed better than DLS based on the MAEs, but it failed to reproduce the linear trend in Layers 13–25 (Fig. 6b).

This occurs, because the penalty term of the $l_2$-smoothness regularization does not strictly hold the average velocity gradient constant, as it is composed of the $l_2$-sum of the $l_2$-norm ($g_z$; Eq. (4)). In contrast, the decrease in precision of the proposed method is relatively suppressed, especially in the layers below the velocity jump (Layers 13–25), by expressing the piecewise linear trend of the true velocity structure via the penalty term, which consists of the $l_1$-sum of the $l_2$-norm (sum of the square roots of $g_z$). We confirmed that the proposed method estimated velocity parameters more stably for each of the tested velocity jumps compared with the conventional methods (Fig. 6d). These results suggest that the proposed method can recover a range of velocity changes (small to large amplitudes) that may be associated with discontinuities.

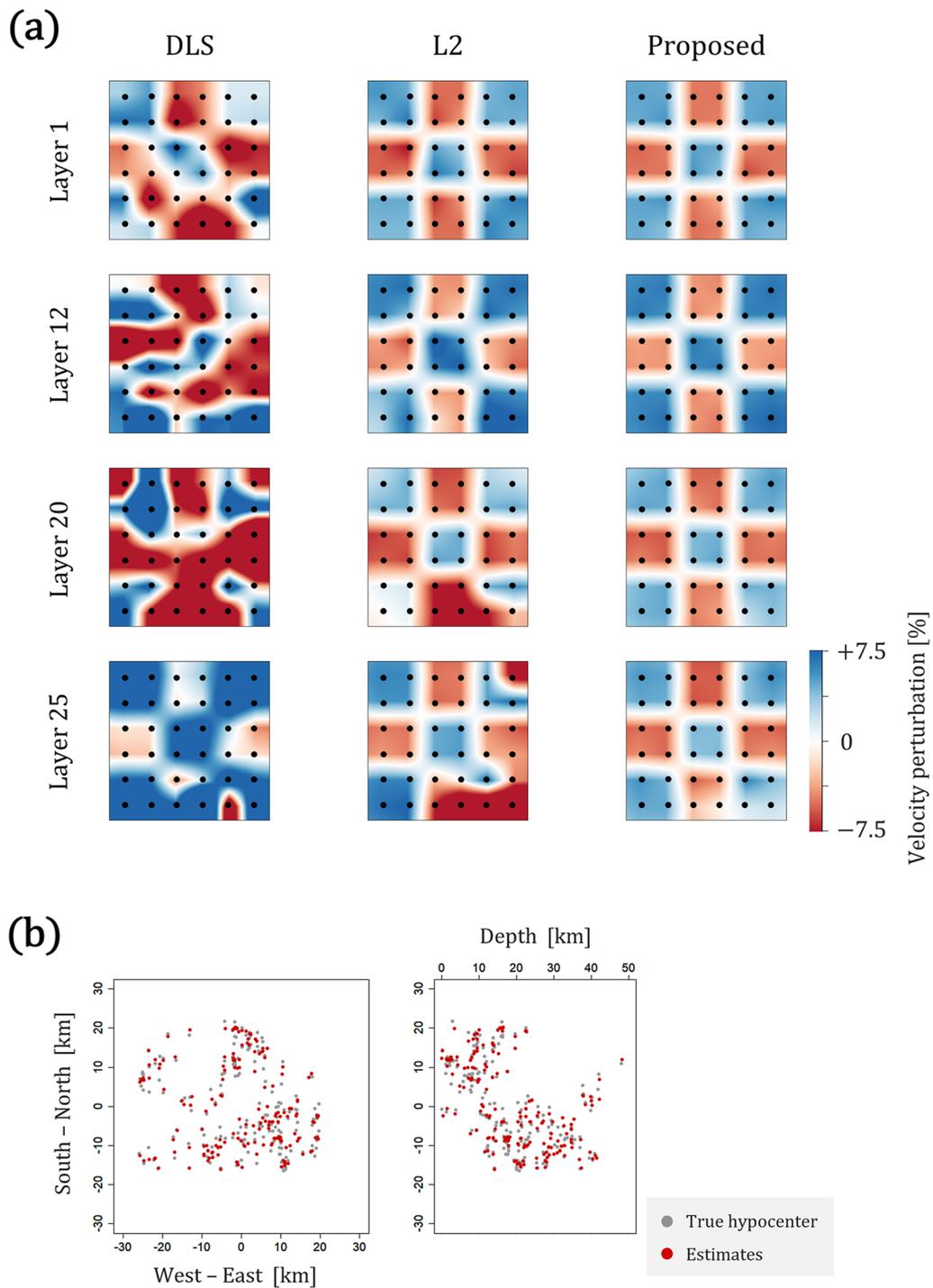

**Fig. 5** **a** Anomalies of the estimated velocities from the baseline velocities in several layers for the estimation methods. Dots represent grid points. **b** Locations of relocated hypocenters in map-view (left) and cross-sectional view along a north–south profile (right). Grey and red circles represent the true (initial) and relocated hypocenters, respectively. East, north, and down directions are positive. The origin of the coordinates is 35.25° N, 138.25° E, and 0 km depth

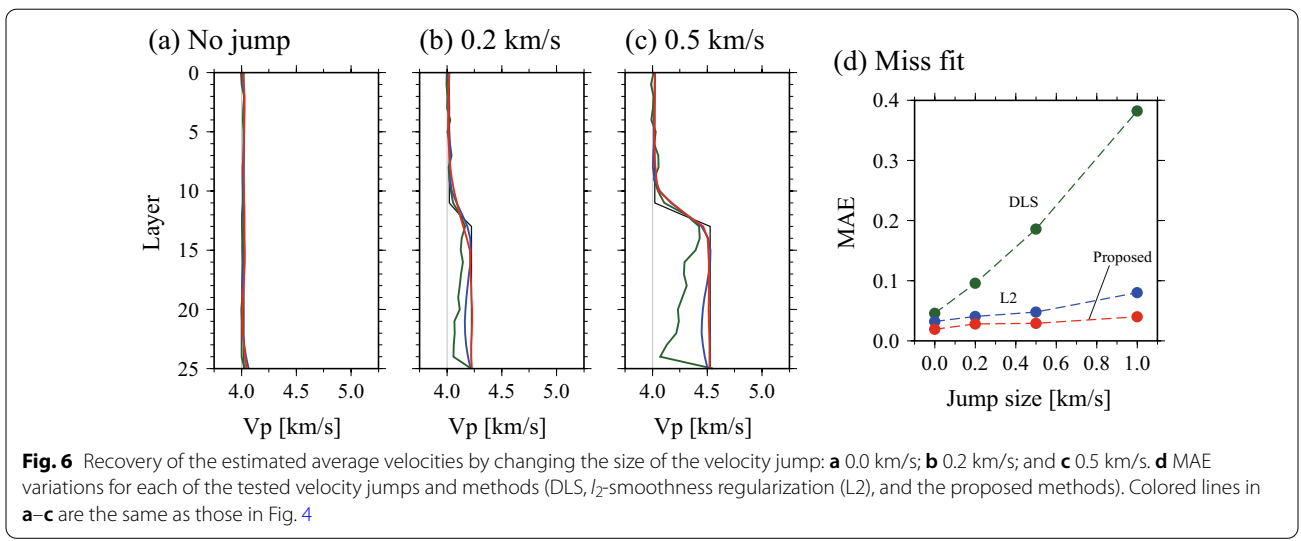

**Fig. 6** Recovery of the estimated average velocities by changing the size of the velocity jump: **a** 0.0 km/s; **b** 0.2 km/s; and **c** 0.5 km/s. **d** MAE variations for each of the tested velocity jumps and methods (DLS, $l_2$-smoothness regularization (L2), and the proposed methods). Colored lines in **a**–**c** are the same as those in Fig. 4

### Initial model dependence

We conducted additional experiments to investigate the influence of the initial model on the estimated velocity structure for each of the estimation method, which may be due to the nonlinearity of the objective function. The main experiment adopted an initial velocity structure of 4.0 km/s at all of the grid points (Fig. 4); our additional experiments tested initial velocities of 4.5 and 5.0 km/s. We configured all of the other settings to be the same as those in the main experiment. The averages of the true and estimated velocities in each layer, and the associated MAEs for different initial velocities are shown in Fig. 7. Note that the proposed method yields the smallest MAEs among the estimation methods for each of the three initial velocities (Fig. 7c), indicating that our structured regularizations provide stable estimations of the velocity structure, regardless of the initial velocity model. It is generally more difficult to estimate the velocity parameters in the deeper layers compared with the shallow layers because of the sparsity of the seismic ray paths at depth. The estimated average velocity more closely reproduced the true velocity in all cases when the initial velocity was set to 5.0 km/s, which is close to the true velocity of the deep layers.

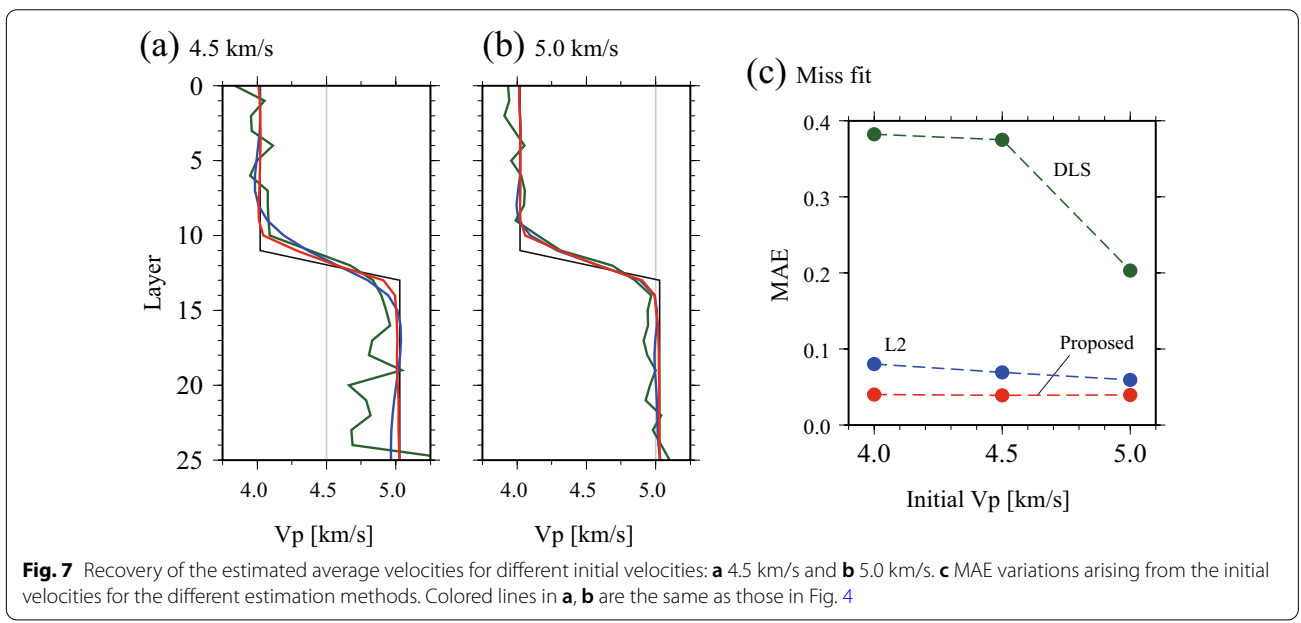

**Fig. 7** Recovery of the estimated average velocities for different initial velocities: **a** 4.5 km/s and **b** 5.0 km/s. **c** MAE variations arising from the initial velocities for the different estimation methods. Colored lines in **a**, **b** are the same as those in Fig. 4

### The relationship between method accuracy and sample size

We investigated the accuracy of each method for different sample sizes (the number of arrival time data). The sample size was controlled by either decreasing or increasing the number of available seismic stations that were analyzed to extract the P-wave arrival times. We used the same amplitude of the velocity jump and initial velocity parameters as those in the main experiment. Figure 8 shows the averages of the true and estimated velocities in each layer, and the MAEs for each sample size. We confirmed that the proposed method performed the best in each of the tested settings based on its MAE values (Fig. 8d). The number of velocity parameters was as large as 936 ($6 \times 6 \times 26$) in our experiments, inevitably making it difficult to estimate the velocity structure without regularization considering the spatial information when the sample size was small (Fig. 8a). Although all methods performed well for a large sample size (Fig. 8b, c), the methods with regularizations showed better accuracies than that of the DLS. The $l_2$-smoothness regularization and proposed methods yielded relatively stable accuracies, even when the number of arrival time data was small, as regularization methods are generally capable of avoiding overfitting and performing well when there are a lot of parameters to estimate (e.g., Negahban et al. 2012; Hastie et al. 2015). Furthermore, the proposed method reproduced the sharp change in the average velocity structure the best among the estimation methods.

Throughout the experiments in the previous and this subsections, $l_2$-smoothness regularization showed a tendency to make biased estimates at depths below the change point (Layers 13–25). In contrast, our proposed method provided less biased and more stable estimates.

### Dipping interface

We assumed that the structure was composed of horizontal (flat) interfaces in the previous experiments, but the interfaces are not always horizontal for more complex structures in Earth. Here, we conducted an additional numerical experiment assuming a dipping interface at crustal depths. We used the same settings as in the main experiment, but assigned a west–east dipping interface in Layers 8–14, shown in the left of Fig. 9, as a true velocity structure. We configured the true average velocity to be piecewise linear: the velocity gradient of the average velocity was constant in Layers 0–7, 8–14, and 15–25, respectively. The number of the observed arrival time data was 4563. Figure 9 also illustrates the west–east vertical cross sections obtained using the three estimation methods (35.43° N). The proposed method recorded the best (smallest) MAE of the estimation methods: the MAE values of DLS, $l_2$-smoothness regularization, and our proposed methods were 0.333, 0.087, and 0.064, respectively. As demonstrated in the previous section and subsections, the proposed method can enhance the flat discontinuities, but on the basis of this experiment, we confirm that it can also be applied to dipping interfaces.

### Application to real seismic data

We applied the proposed method to real seismic data, using seismic waveforms from 211 earthquakes that were observed by the high-sensitivity seismograph network in

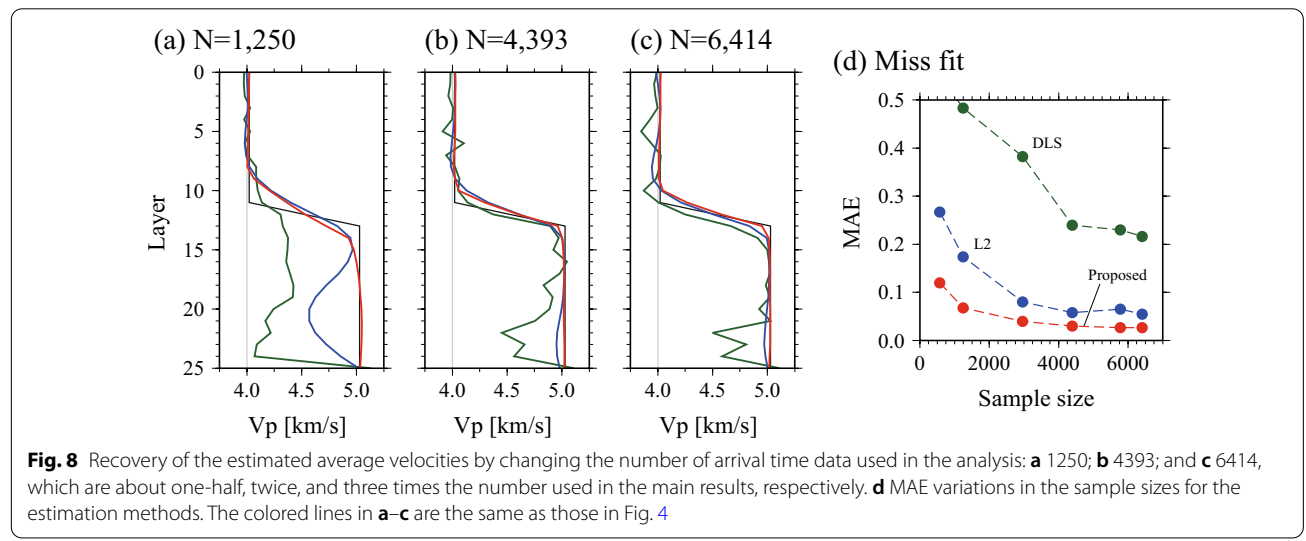

**Fig. 8** Recovery of the estimated average velocities by changing the number of arrival time data used in the analysis: **a** 1250; **b** 4393; and **c** 6414, which are about one-half, twice, and three times the number used in the main results, respectively. **d** MAE variations in the sample sizes for the estimation methods. The colored lines in **a**–**c** are the same as those in Fig. 4

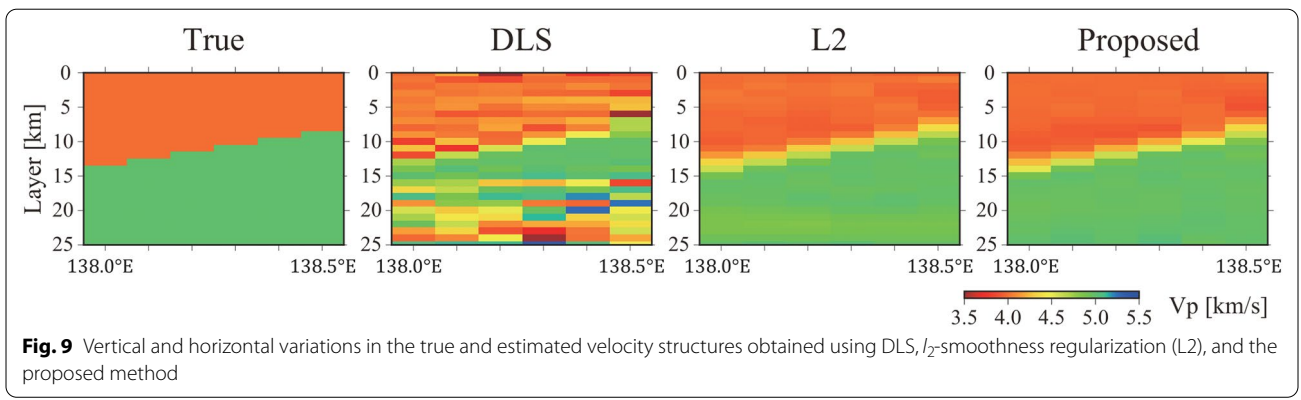

**Fig. 9** Vertical and horizontal variations in the true and estimated velocity structures obtained using DLS, $l_2$-smoothness regularization (L2), and the proposed method

Japan (Hi-net, National Research Institute for Earth Science and Disaster Resilience 2019) during the 2005–2014 period. We used 2042 P-wave arrival times from the waveforms, and divided the arrival times into 1701 training data and 341 validation data for cross-validation. The target region of this experiment is shown in Fig. 2. We employed the same grid points as those in the synthetic test. We set a velocity of 6.0 km/s at all of the grid points in the study area for the initial velocity model, and fixed the JMA2001 1-D velocity model (Ueno et al. 2002) values, which have been commonly employed for routine hypocenter determinations throughout Japan, at the outer points (outside the target region).

The resultant P-wave velocity–depth profiles for the methods are shown in Fig. 10a–c illustrate the vertical cross-sectional variations. The proposed method estimated a notable change of averaged velocity in the target region, with monotonically increasing averaged velocities to approximately 16 km depth (Layer 16) and a nearly constant velocity at greater depths (red line in Fig. 10a). From Fig. 10, it can be seen that there is a change in velocity gradient at the depth around 16 km (arrows in Fig. 10b, c). The obtained average velocities at depths greater than 16-km depth were approximately 6.71 km/s, coinciding with those determined by the reflection and wide-angle refraction survey (Iidaka et al. 2003). Since the Conrad discontinuity in the target region has been imaged at approximately 15–20 km depth (e.g., Iidaka et al. 2003; Katsumata 2010), we interpreted that the change in average velocity gradient at depths of around 16 km could be related to this discontinuity.

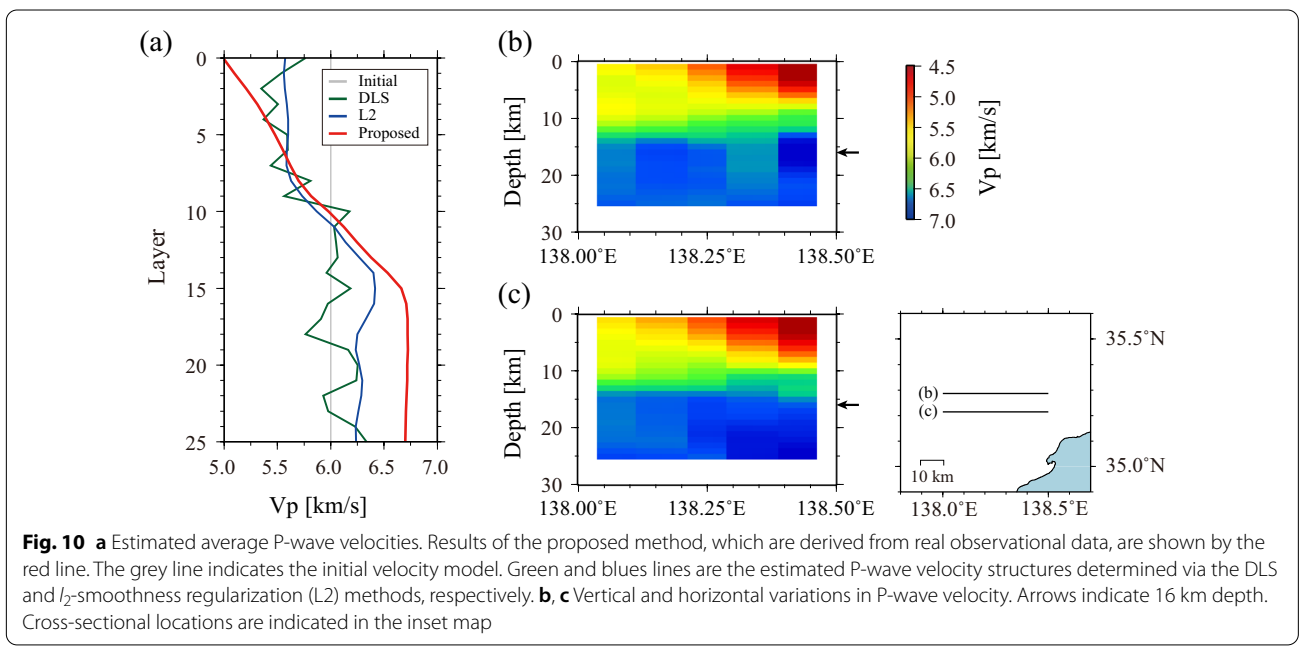

**Fig. 10** **a** Estimated average P-wave velocities. Results of the proposed method, which are derived from real observational data, are shown by the red line. The grey line indicates the initial velocity model. Green and blues lines are the estimated P-wave velocity structures determined via the DLS and $l_2$-smoothness regularization (L2) methods, respectively. **b**, **c** Vertical and horizontal variations in P-wave velocity. Arrows indicate 16 km depth. Cross-sectional locations are indicated in the inset map

The proposed method estimated the eastward dipping interface in the shallower part (Fig. 10b, c). The obtained mean and standard deviation of P-wave velocity in the west of the 0–5 km depth were 5.66 and 0.20 km/s, that is comparable with those determined by Iidaka et al. (2003) of which the survey lines crossed the west of the target area. Meanwhile, obtained mean and standard deviation of P-wave velocity at the east of the 0–5 km depth were 4.83 and 0.54 km/s, respectively. Similar near-surface low-velocity zones has been imaged by Matsubara et al. (2019) around the east of the target area, supporting our results. Since we took the depth–average including the low-velocity region, the obtained average velocity was gradually increased down to the depths of about 16 km and thus the change of average velocity at the deeper potion would become somewhat continuously.

Therefore, the applicability of the proposed method in elucidating sharp velocity discontinuities is validated by its success in detecting the structural change, which is defined by this sudden change in the velocity gradient within the target region. There were large fluctuations in the DLS average velocities (green line in Fig. 10a), whereas the $l_2$-smoothness regularization vertical fluctuations were smoothed (blue line in Fig. 10a). The average velocity gradient of the $l_2$-smoothness regularization shows some changes in the Layers 13–16 range, as well as the proposed method. However, the P-wave velocity obtained by the $l_2$-smoothness regularization method was 6.28 km/s at the depths of greater than 16 km, which was clearly smaller than those retrieved by the reflection and wide-angle refraction survey, 6.6–6.8 km/s (Iidaka et al. 2003), and the proposed method, 6.71 km/s. The small number of arrival time data used here can cause underestimations of average velocities in the $l_2$-smoothness regularization method.

The regularization parameters for the proposed methods were ($\lambda_{ver} = 0.45, \lambda_{hor} = 0.95$). RMSE values for each pair of regularization parameters ($\lambda_{ver}, \lambda_{hor}$) are represented by a heat map (Fig. 11a, b). We also show the estimated average velocities and values of $g_z$ (Eq. (4)) for each layer for some pairs of ($\lambda_{ver}, \lambda_{hor}$), in Fig. 11c–e. The RMSE for optimal regularization parameter was 0.17 s. When using values of $\lambda_{ver}$ and $\lambda_{hor}$ that are too small, the corresponding estimation procedure is similar to that of DLS, and thus it becomes difficult to adapt to a sharp change in the velocity structure (Fig. 11c). In contrast, when using values of $\lambda_{ver}$ and $\lambda_{hor}$ that are too large, variations in velocity gradients and adjacent velocity parameters are suppressed excessively, and the resultant velocity structure tends to be too smooth (Fig. 11e).

These results, which are obtained from real seismic data, suggest that the proposed method can stably detect the true depth of the velocity discontinuity, even when the number of available observational data is small. Later reflected and/or converted waves have conventionally been used to investigate the depths of various velocity discontinuities, such as the Conrad and Moho discontinuities, and the subducting plate interface (e.g., Matsuzawa et al. 1986; Zhao et al. 1997). However, there are cases, where such later waves are identified only in a limited number of ray paths of earthquake–station pairs, unlike direct P and S waves that are commonly and widely observed from numerous earthquakes. As mentioned in the previous section, the estimated accuracy of our method improves with increasing sample size, as with conventional methods; thus, available later arrival data will be useful for improving the accuracy of estimation of the proposed method. A significant advantage of the proposed method is that it can estimate velocity structure robustly, even in cases where there is only a small number of data, by employing sparse regularization. The proposed method will improve the detection of velocity discontinuities considerably and refine imaging in regions, where later waves are not widely observed.

## Conclusions

We introduced a nonlinear inversion method with structured regularization to image the crustal structure of Earth. Our proposed LET method simultaneously estimated both smooth trends and sharp changes in crustal velocity structure, both of which are expected in Earth's interior, by combining two types of penalty terms that are added to the vertical and horizontal directions of the model space. We employed a vertical-direction penalty term that consisted of the second-order differences in the depth-dependent velocity parameters to detect a velocity discontinuity, thereby highlighting the ability to image sharp velocity changes in the vertical direction. This vertical-direction penalty term works on the depth-averaged velocity values, and takes the form of the $l_1$-sum of the $l_2$-norm. This penalty enables to reproduce piecewise linear trends in the velocity changes at depth, and image the sharp structural changes. We used a horizontal-direction penalty term that consisted of first-order differences of the velocities that were based on the $l_2$-norms. This horizontal-direction penalty term smooths velocity fluctuations.

We compared the imaging capability of the proposed method with conventional LET approaches, the damped least-squares and $l_2$-based regularization methods, via synthetic data experiments to verify the performance of the proposed method. Accordingly, we confirmed that the proposed method can adequately reproduce both of sharp and gradual velocity changes. We also demonstrated that the proposed method is stable against variations in the amplitude of velocity jump, initial velocity structure, and sample size, and that it has the ability to accommodate dipping structural changes in the crust. Furthermore, we applied the proposed method to real seismic data from central Japan, and successfully imaged a distinct velocity gradient change at approximately 16 km depth. Therefore, the proposed method can improve the detectability of horizontal and dipping interfaces using arrival time data. Our proposed method automatically detects the existence (or nonexistence) of discontinuities, because it does not require prior information regarding the velocity discontinuity. Results of the synthetic tests and the real data analysis highlighted the importance of sparse regularization to better estimate the subsurface velocity structure, and suggested that we can improve the imaging in the framework of the earthquake tomography for existing seismic data by combining appropriately the structured regularizations.

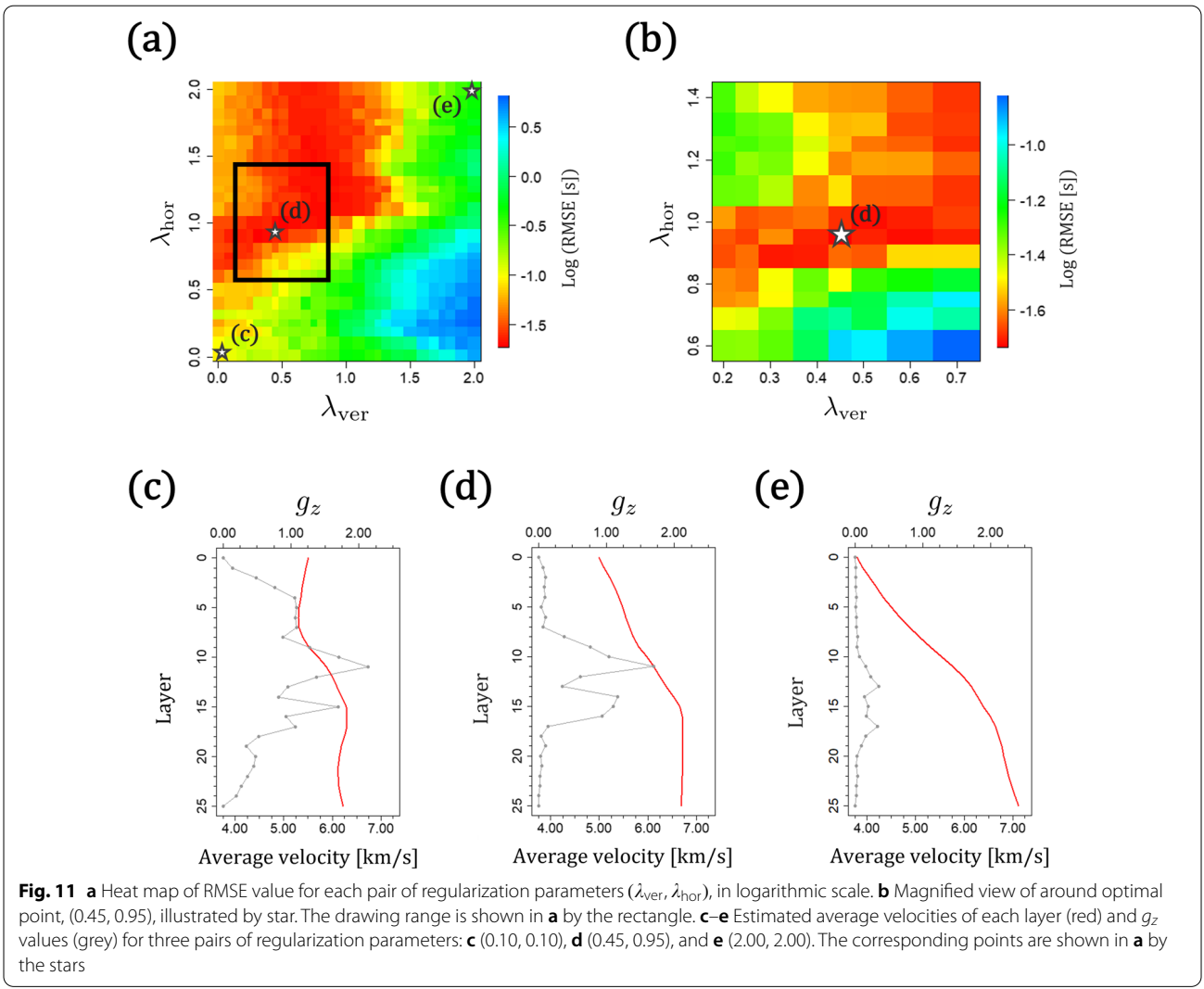

**Fig. 11** **a** Heat map of RMSE value for each pair of regularization parameters ($\lambda_{\text{ver}}$, $\lambda_{\text{hor}}$), in logarithmic scale. **b** Magnified view of around optimal point, (0.45, 0.95), illustrated by star. The drawing range is shown in **a** by the rectangle. **c–e** Estimated average velocities of each layer (red) and $g_z$ values (grey) for three pairs of regularization parameters: **c** (0.10, 0.10), **d** (0.45, 0.95), and **e** (2.00, 2.00). The corresponding points are shown in **a** by the stars

## Appendix: An optimization with ADMM

Here, we introduce the estimation procedure of our proposed method. The RSS (denoted by $R(v, h)$ in this section) and the damping term in the objective function in Eq. (1) depend on hypocentral parameters $h$ other than velocity parameters $v$. Thus, we first approximate $R(v, h) + D(v, h)$ by a quadratic form. We then separate the quadratic form into an additive form of two terms depending on only the velocity parameters (hereafter, denoted by $l(v)$) and hypocentral parameters, by utilizing QR decomposition. Thus, the objective function of the proposed method is

$$f(\nu) = l(\nu) + \lambda_{\text{ver}} \Omega_{\text{ver}}(\nu) + \frac{1}{2} \lambda_{\text{hor}} \Omega_{\text{hor}}(\nu).$$

Now, we obtain the following optimization problem via the duality principle:

$$\text{minimize}_{\nu,w} \left[ \frac{1}{\lambda_{\text{ver}}} \left\{ l(\nu) + \frac{\lambda_{\text{hor}}}{2} \Omega_{\text{hor}}(\nu) \right\} + \sum_z ||w_z||_2 \right],$$

subject to $A_z \nu - w_z = 0$ $(z = 2, \ldots, n_z - 1)$,

where $|| \cdot ||_2$ represents the $l_2$-norm

$$w = (w_2, \ldots, w_{n_z-1})^T,$$
$$w_z = u_{z-1} - 2u_z + u_{z+1} \ (z = 2, \ldots, n_z - 1),$$

$$u_z = (\nu_{1,1,z}, \ldots, \nu_{1,n_y,z}, \nu_{2,1,z}, \ldots, \nu_{2,n_y,z}, \ldots,$$
$$\nu_{n_x,1,z}, \ldots, \nu_{n_x,n_y,z})^T \quad (z = 1, \ldots, n_z),$$

where $A_z$ is a matrix satisfying $A_z \nu = w_z$ $(z = 2, \ldots, n_z - 1)$, and $n_x, n_y$, and $n_z$ are the numbers of grid points in $x$-, $y$-, and $z$-directions, respectively. We then apply the augmented Lagrange multiplier method, and consider the following function:

$$L_\eta(\nu, w, \alpha) = \frac{1}{\lambda_{\text{ver}}} \left\{ l(\nu) + \frac{\lambda_{\text{hor}}}{2} \Omega_{\text{hor}}(\nu) \right\}$$
$$+ \sum_z ||w_z||_2$$
$$+ \sum_z \left\{ \alpha_z^T (A_z \nu - w_z) \right.$$
$$+ \left. \frac{\eta}{2} ||A_z \nu - w_z||_2^2 \right\}.$$

We obtain estimates by applying the ADMM (Glowinski and Marroco 1975; Gabay and Mercier 1976), a widely used algorithm that is well suited for solving distributed convex optimization problems (e.g., Boyd et al. 2011; Li and Harris 2018):

$$\nu^{(t+1)} = \text{argmin}_\nu L_\eta \left( \nu, w^{(t)}, \alpha^{(t)} \right)$$
$$= \text{argmin}_\nu \left[ \frac{1}{\lambda_{\text{ver}}} \left\{ l(\nu) + \frac{\lambda_{\text{hor}}}{2} \Omega_{\text{hor}}(\nu) \right\} \right.$$
$$+ \left. \frac{\eta}{2} \sum_z ||A_z \nu - w_z^{(t)} + \frac{1}{\eta} \alpha_z^{(t)}||_2^2 \right],$$

$$w^{(t+1)} = \text{argmin}_w L_\eta \left( \nu^{(t+1)}, w, \alpha^{(t)} \right)$$
$$= \text{argmin}_w \sum_z \left\{ ||w_z||_2 + \frac{\eta}{2} \sum_z \right.$$
$$\left. ||A_z \nu^{(t+1)} - w_z + \frac{1}{\eta} \alpha_z^{(t)}||_2^2 \right\},$$

$$w_z^{(t+1)} = \text{prox}_{\frac{1}{\eta}||\cdot||_2} \left( A_z \nu^{(t+1)} + \frac{1}{\eta} \alpha_z^{(t)} \right),$$

$$\alpha_z^{(t+1)} = \alpha_z^{(t)} + \eta \left( A_z \nu^{(t+1)} - w_z^{(t+1)} \right)$$
$$(z = 2, \ldots, n_z - 1),$$

where "prox" is the proximal operator. We note that the minimization function for $\nu$ is nonlinear and nonconvex, and that it takes an additive form of the sums of squares via iterative quadratic approximation. Therefore, we obtain the optimal value using the Levenberg–Marquardt method (e.g., Levenberg 1944; Marquardt 1963; Gavin 2013).

## Comparison of the proposed method with various regularization methods

In this section, we additionally compare our proposed method with other regularization methods; the Laplacian regularization and two sparse regularization methods. The Laplacian-based regularization has been used in LET (e.g., Lees and Crosson 1989; Zhang et al. 1998; Moran et al. 1999). The penalty term, $P(\nu)$ in Eq. (1), is given as

$$P^{\text{Lap}}(\nu) = \lambda_{\text{Lap}} \sum_{x,y,z} ||\Delta \nu_{x,y,z}||_2^2,$$

where $\lambda_{\text{Lap}}$ is a non-negative regularization parameter. Here, we refer to this Laplacian regularization method as "Lap" for notational simplicity. Moreover, in earthquake tomography, some studies have applied sparse regularization methods via $l_1$-norm (e.g., Zhang et al. 2014). As described in the Introduction, $l_1$-type regularizations yield sparse estimation, by shrinking less important features (penalized elements) to zero. We here employ a gridwise regularization using $l_1$-norm, that penalizes the first-order differences among adjacent grid points. We term this method "L1first" for notational simplicity. In L1first, $P(\nu)$ is given as

$$P^{\text{L1first}}(\nu) = \lambda_{\text{L1first}} \sum_{x,y,z} \sum_{(x',y',z') \sim (x,y,z)} |\nu_{x,y,z} - \nu_{x',y',z'}|,$$

where $\lambda_{\text{L1first}}$ is a non-negative regularization parameter. We also examine another regularization method that penalizes the second-order differences among adjacent grid points via $l_1$-norm, referred to as "L1second". $P(\nu)$ in L1second is given as

$$P^{\text{L1second}}(\nu) = \lambda_{\text{L1second}} \sum_{x,y,z} ||\Delta \nu_{x,y,z}||_1,$$

where $|| \cdot ||_1$ represents the $l_1$-norm, and $\lambda_{\text{L1second}}$ is a non-negative regularization parameter. The relationship between L1second and Lap is similar to that between the proposed method and $l_2$-smoothness regularization;

L1second reduces penalized elements (second-order differences among velocity parameters of adjacent grid points) to zero exactly, whereas Lap does not. Note that, Lap, L1first, and L1second employs the same types of penalty terms for both the vertical and horizontal directions, respectively, unlike $l_2$-smoothness regularization and our proposed method, which have different types of penalty for the vertical and horizontal directions by taking the characteristics of Earth's seismic velocity structure into consideration.

### Results of the main experiment using Laplacian regularization and sparse regularizations via $l_1$-norm

We verified the accuracies of estimation of the abovementioned methods in the main experiment in the section of the main text entitled "Numerical experiment". Additional file 1: Fig. S1 presents the estimated average velocities and horizontal-direction anomalies from the baseline velocities for the estimation methods. The values of MAE of Lap, L1first, and L1second were 0.185, 0.161, and 0.135, respectively. Although the three methods outperformed DLS in terms of MAE (the MAE value of DLS was 0.383), none of them recorded better (smaller) MAE values than those of $l_2$-smoothness regularization (0.080) and our proposed method (0.040). From Additional file 1: Fig. S1, we can see that Lap produced estimates far from the true structure at many grid points, and L1first smoothed the checkerboard anomalies too much by reducing the variation in velocity parameters among adjacent grid points. L1second reproduced the checkerboard anomalies better than Lap and L1first, but it estimated opposite polarities of velocity anomalies at some grid points (e.g., Layers 1 and 20 in Additional file 1: Fig. S1).

### Synthetic test for a three-dimensional checkerboard pattern

In the main experiment, we produced the checkerboard pattern only in the horizontal direction to verify the performance of the horizontal-direction penalty terms. Here, we assign the checkerboard pattern of velocity perturbations to the uniform velocity structure in both the horizontal and vertical directions. The number of arrival time data was 4,056. In this experiment, we used $6 \times 6 \times 24$ grid points, and generated a 3-D checkerboard velocity model from the uniform velocity structure of 4.0 km/s using perturbations of $\pm 5\%$: we reversed the positive/negative number of the velocity perturbations every grid point in the horizontal direction, and every four grid points (layers) in the vertical direction. Additional file 1: Fig. S2 illustrates the EW depth-profile of the true and recovered velocity structures ($35.21°$ N).

Although some velocity anomalies in deep layers were not well reproduced, the accuracy of estimation of the proposed method showed the best score index in this experiment: the values of MAE for DLS, Lap, L1first, L1second, $l_2$-smoothness regularization, and our proposed method were 0.084, 0.075, 0.085, 0.088, 0.071, and 0.070, respectively.

### The case for which there are multiple horizontal layers at depth

As the performances of DLS and Lap were worse than those of the other methods, we focus on comparison of L1first, L1second, and $l_2$-smoothness regularization (abbreviated as "L2") with our proposed method, in this and the next subsection.

In this subsection, we describe an experiment assuming the case for which there are multiple horizontal layers at depth. The number of arrival time data was 5025. In this experiment, we assumed that all values of the velocity parameter were uniform in each layer. The other settings were the same as those in the main experiment. Additional file 1: Fig. S3 shows the true average velocity structure and the results estimated by each method. The MAE values of L1first, L1second, $l_2$-smoothness regularization, and our proposed method were 0.042, 0.045, 0.031, and 0.015, respectively. Results of this experiment suggest that the proposed method can closely reproduce the velocity profile containing multiple horizontal layers.

### The case for which there is a high-velocity zone in the target region

We conducted an experiment assuming a high velocity anomaly of 5.0 km/s embedded in the homogeneous velocity medium of 4.0 km/s to verify the robustness of estimation methods. The number of arrival time data was 4529. The other settings were the same as the main experiment. The upper-left part of Additional file 1: Fig. S4 shows the vertical and horizontal variations (south–north profile) in true velocity structure ($138.21°$ E). As the earthquake distribution is biased toward the north part of the target region (also see Fig. 2), many ray-paths pass through the high-velocity zone; in contrast, relatively few ray paths pass through the south part. Additional file 1: Fig. S4 also illustrates the estimated structure by each method. MAE values of L1first, L1second, $l_2$-smoothness regularization, and our proposed methods were 0.120, 0.124, 0.115, and 0.088, respectively. Although the accuracy of estimation around the structural boundary in this setting was somewhat worse than in the other experiments due to the biased distribution of ray paths, the proposed method performed the most robustly with respect to the high-velocity anomaly.

## Abbreviations

ADMM: Alternating direction method of multipliers; DLS: Damped least-square; Hi-net: High-sensitivity seismograph network; JMA: Japan Meteorological Agency; LET: Local earthquake tomography; MAE: Mean absolute error; NIED: National Research Institute for Earth Science and Disaster Resilience; RMSE: Root mean square error; RSS: Residual sum of squares.

## Supplementary Information

The online version contains supplementary material available at https://doi.org/10.1186/s40623-022-01600-x.

**Additional file 1.** Additional figures.

## Acknowledgements

We used data recorded on seismograph networks operated by the Japan Meteorological Agency and the National Research Institute for Earth Science and Disaster Resilience (Hi-net). GMT software package (Wessel and Smith 1998) and R (R Core Team 2020) were used for creating figures. We deeply thank the AE and two reviewers for their valuable comments and suggestions.

## Authors' contributions

YY conceptualized this study, supported by SK, KY, and FK. YY and SK carried out the analyses, and SK validated the results. SK drafted the manuscript, supported by KY and TS. TS and AK contributed to the implications of the results. FK and AK supervised this work. All authors read and approved the final manuscript.

## Funding

This work was supported by JST CREST Grant Number JPMJCR1763, MEXT Grant Number JPJ010217, and JSPS KAKENHI Grant Numbers JP20K19753 and JP21H05205.

## Availability of data and materials

The data sets analysed during the current study are available from the Japan Meteorological Agency (JMA) at https://www.data.jma.go.jp/svd/eqev/data/bulletin/deck_e.html.

## Declarations

### Competing interests

The authors declare that they have no competing interests.

### Author details

[1]Graduate School of Information Science and Technology, The University of Tokyo, Tokyo, Japan. [2]The Institute of Statistical Mathematics, Tokyo, Japan. [3]Geological Survey of Japan, National Institute of Advanced Industrial Science and Technology (AIST), Ibaraki, Japan. [4]Earthquake Research Institute, The University of Tokyo, Tokyo, Japan.

## Publisher's Note